\documentclass
[superscriptaddress,secnumarabic,amssymb,amsmath,nobibnotes,aps,prd,showkeys,showpacs,nofootinbib]{revtex4}%
\usepackage{graphicx}
\usepackage{epsf}
\usepackage{bm}
\usepackage{amsmath}
\usepackage{amsfonts}
\usepackage{amssymb}
\usepackage{epstopdf}%
\providecommand{\U}[1]{\protect\rule{.1in}{.1in}}
\newcommand{\be}{\begin{equation}}
\newcommand{\ee}{\end{equation}}

\newcommand{\mincir}{\raise
-3.truept\hbox{\rlap{\hbox{$\sim$}}\raise4.truept\hbox{$<$}\ }}
\newcommand{\magcir}{\raise
-3.truept\hbox{\rlap{\hbox{$\sim$}}\raise4.truept\hbox{$>$}\ }}

\begin{document}
\title{ Wheeler-DeWitt equation  and Lie symmetries in Bianchi scalar-field cosmology}
\author{A. Paliathanasis}
\email{anpaliat@phys.uoa.gr}
\affiliation{Instituto de Ciencias F\'{\i}sicas y Matem\'{a}ticas, Universidad Austral de
Chile, Valdivia, Chile.}
\author{L. Karpathopoulos}
\email{lkarpathopoulos@gmail.com}
\affiliation{Faculty of Physics, Department of Astronomy-Astrophysics-Mechanics University
of Athens, Panepistemiopolis, Athens 157 83, Greece.}
\author{A. Wojnar}
\email{aneta.wojnar@ift.uni.wroc.pl}
\affiliation{Institute for Theoretical Physics, pl. M. Borna 9, 50-204, Wroclaw, Poland}
\affiliation{Dipartimento di Fisica, Universita' di Napoli Federico II, Complesso
Universitario di Monte S. Angelo, Via Cinthia, 9, I-80126 Naples, Italy.}
\affiliation{Istituto Nazionale di Fisica Nucleare (INFN) Sez. di Napoli, Complesso
Universitario di Monte S. Angelo, Via Cinthia, 9, I-80126 Naples, Italy.}
\author{S. Capozziello}
\email{capozziello@na.infn.it}
\affiliation{Dipartimento di Fisica, Universita' di Napoli Federico II, Complesso
Universitario di Monte S. Angelo, Via Cinthia, 9, I-80126 Naples, Italy.}
\affiliation{Istituto Nazionale di Fisica Nucleare (INFN) Sez. di Napoli, Complesso
Universitario di Monte S. Angelo, Via Cinthia, 9, I-80126 Naples, Italy.}
\affiliation{Gran Sasso Science Institue (INFN), Viale F. Crispi 7, I-67100, L' Aquila, Italy.}

\begin{abstract}
Lie symmetries are discussed for the Wheeler-De Witt equation in Bianchi Class A cosmologies. In particular, we consider General Relativity, minimally coupled scalar field gravity and Hybrid Gravity as paradigmatic examples of the approach. 
Several invariant solutions are determined and classified according to the form of the scalar field potential.  The approach gives rise to a suitable method to select classical solutions and it is  based on the first principle of the existence of symmetries.

\end{abstract}
\maketitle

\section{Introduction}

Nowadays astronomical observations have shown that if we consider our Universe
on a large scale, its visible structure is accelerating, homogeneous and isotropic and, essentially,  filled 
with pressureless  dust. The simplest cosmic model which
describes  a Universe with  the above properties is the Friedmann-Lema\^itre-Robertson-Walker (FLRW) model \cite{kolb}. The evolution of the Universe from the
radiation dominant epoch till the present cosmic acceleration can be
well-explained by the FLRW model with a cosmological constant (the so-called
$\Lambda$CDM model). However, it fails if one tries to address  the whole early and late
history of the Universe starting from the origin and the inflation epoch where quantum effects should
be  taken into account.

Anisotropies observed in the cosmic microwave background (CMB) are small
enough to suggest that anisotropic models of spacetimes become isotropic ones
by evolving in  time \cite{Mis69, szydl, russ}. One can expect that pre-inflationary
anisotropies played an important role (for example they can be responsible for
the coupling between the gravitational field and the inflaton field) so if
inflationary models are considered, one should understand the dynamics of
anisotropies. Models describing anisotropic but homogeneous universes are the so-called 
{\it Bianchi cosmologies}. They can be considered in standard 
General Relativity and in its extensions containing scalar fields.

In this paper we will consider the Lie symmetries of the Wheeler-DeWitt equation
(WDW) in General Relativity and in scalar field cosmology assuming Bianchi spatially homogeneous
spacetimes. We will use the Lie symmetries in order to define the
unknown potential and derive exact solutions for the WDW equation and for the
field equations. Symmetries are considered to play a central role in physical
problems because they provide first integrals which can be utilized in order
to simplify a given system of differential equations and thus to determine the
integrability of the system. Indeed, in \cite{GRG,AnIJGMMP} it has been shown
that the Lie symmetries of a dynamical system are related to the geometry of
the underlying space where dynamics occurs.

The application of symmetries in gravitational theories  is an important tool which could
lead to exact solutions of the field equations. In particular, the symmetries
which can be used are the  Noether symmetries of the Lagrangian of
the field equations and they have been applied in several  models  such as  scalar tensor cosmology
\cite{Cap93deR,CapP09,AAslam,YiZhang,Basilakos,Kukum,BVakili,deSouza}, $f(R)$ gravity  and higher order theories of gravity
\cite{PalFR,VakF,dimakisT,CapDF,CamL,Ermakov}, spherically symmetric
spacetimes \cite{Cap2012,AnFT,TchRN,TchSA} and many others.
\cite{SBasil,HWei,KucukE,Hybrid}. The application of the Noether symmetries in
Bianchi spatially homogeneous spacetimes can be found in
\cite{GRG,CapB,Cotsakis,VakB}. According to this results one can deal with the so-called {\it Noether Symmetry Approach}.

In this work we will not apply the Noether symmetries of the field
equations but the Lie symmetries of the WDW equation. It has been proved
in \cite{AnIJGMMP} that the Lie symmetries of the WDW equation could form a
greater Lie algebra than the Noether symmetries of the Lagrangian of the field
equations. Hence, it is possible to determine new cases where the field
equations are integrable. This method was applied in a scalar
tensor cosmological models adopting a
FLRW geometry with a perfect fluid and new integrable models,  
cosmologically viable, raised \cite{bar1,bar2}. Recently a similar method has been applied
to some axisymmetric quantum cosmologies with scalar fields \cite{manto}.

The layout of the paper is the following. 
In Sec. \ref{Bianchi} we give the basic definition of the Bianchi
classification while in Sec. \ref{BianchiWDW} we study the Lie symmetries of
the WDW equation of class A Bianchi spacetimes in the case of General
Relativity. In  Sec. \ref{BianchiWDWI}, the previous results are applied
in order to reduce the WDW equation by using the Lie invariants and determine
invariant solutions. Secs. \ref{BianchiSF} and \ref{BianchiSFWDW} are devoted to  the same analysis in the case of a minimally coupled scalar tensor  gravity  and we
use the Lie symmetries in order to determine the field potential by using the Lie
symmetries of the WDW equation as a geometric criterion. We show that, in
scalar tensor  cosmology, there exist invariant solutions of the WDW equation:  in
Bianchi I spacetime,  for constant potential and for exponential 
potential and, in Bianchi II spacetime, for a kination  scalar field.  For
convenience of the reader,  we present the Lie symmetry classification for each
model in tables. Furthermore in Sec. \ref{hybrid}, we show how these results
are related with the so-called Hybrid Gravity and conformal transformations.  Finally, in  Sec. \ref{conclusion},  we
discuss our results and  draw  conclusions. 
Appendix
\ref{basictools} completes our presentation. Here the basic theory of  Lie
symmetries is briefly discussed.

\section{The Class A of Bianchi spacetimes}

\label{Bianchi}

The class of Bianchi spatially homogeneous cosmologies contains several important
cosmological models which  have been used for the discussion of anisotropies
of primordial Universe and for  its evolution towards the observed isotropy of
the present epoch \cite{Mis69,DemB,jacobs2,narlikan,Barrow}. In these models,  the
spacetime manifold is foliated along the time axis, with three dimensional
homogeneous hypersurfaces which admit a group of motions $G_{3}$. Bianchi
classified all three dimensional real Lie algebras and has shown that there
are nine possible $G_{3}$ groups. This results in nine types of Bianchi
spatially homogeneous spacetimes. The importance of Bianchi cosmological
models is that, in these models,  the physical variables depend on the time
only, reducing the Einstein and other governing equations to ordinary
differential equations.

In the $(3+1)$ decomposition of spacetime manifold (Arnowitt-Deser-Misner  (ADM) formalism), the line
element of the Bianchi models can be written in the following form
\cite{BRyan,Misner}
\begin{equation}
ds^{2}=-\frac{1}{N(t)^{2}}dt^{2}+\bar{g}_{ij}(t)\omega^{i}\otimes\omega
^{j},\label{B2}%
\end{equation}
where $N(t)$ is the lapse function and $\{\omega^{i}\}$ denotes the canonical
basis $1$-forms satisfying the Lie algebra
\begin{equation}
d\omega^{i}=C_{jk}^{i}\omega^{j}\wedge\omega^{k}\label{B2.2}%
\end{equation}
where $C_{jk}^{i}$ are the structure constants of the algebra. The spatial
metric $\bar{g}_{ij}$ is diagonal (following the notation of \cite{GRG,CapB} and
references therein) and can be factorized as follows
\begin{equation}
g_{ij}(t)=e^{2\lambda(t)}e^{-2\beta_{ij}(t)}\label{B2.3}%
\end{equation}
where $e^{\lambda\left(  t\right)  }$ is the scale factor of the universe and
the matrix $\beta_{ij}$ is diagonal and  traceless. The matrix $\beta_{ij}$
depends on two independent quantities $\beta_{1},$ $\beta_{2}~$\ which 
are called the anisotropy parameters. The matrix $\beta_{ij}$ can be selected as\footnote{These are the Misner variables \cite{Misner}.}
\begin{equation}
\beta_{ij}=diag\left(  \beta_{1},-\frac{1}{2}\beta_{1}+\frac{\sqrt{3}}{2}%
\beta_{2},-\frac{1}{2}\beta_{1}-\frac{\sqrt{3}}{2}\beta_{2}\right)
.\label{B2.4}%
\end{equation}
and, in these variables, it is  $\sqrt{\bar{g}}=e^{3\lambda}$.
The structure constants of the Lie algebra $G_{3}$ can be expressed in terms
of a three dimensional vector field $a^{i}$ and a symmetric $3\times3$ tensor
$m^{ij}$ as follow \cite{MacCallumn}
\begin{equation}
C_{jk}^{i}=\epsilon_{jks}m^{si}+\delta_{k}^{i}a_{j}-\delta_{j}^{i}%
a_{k},\label{B2.5}%
\end{equation}
and the Bianchi models are grouped in two classes: class A for $a^{i}=0$ and
Class B for $a^{i}\neq0$. Each class is divided into several types according
to the rank and the signature of the tensor $m^{ij}$.  Specifically, the Bianchi models are
divided into two subclasses A ($a_{i}=0$) and B ($a_{i}\neq0$)  containing
Bianchi types corresponding to the form of the metric $m^{ij}$. In this paper,
we are interested in the class A models for which there exists a Lagrangian of
the field equations.

For the line element (\ref{B2}) with the definitions (\ref{B2.3}) and (\ref{B2.4}), the Ricci
scalar of the Bianchi class A  spacetimes can be written as %
\begin{equation}
R=R_{\left(  4\right)  }+R^{\ast}\label{B4a}%
\end{equation}
where
\begin{equation}
R_{\left(  4\right)  }=\frac{3}{2}N\left(  4N\ddot{\lambda}+4\dot{N}%
\dot{\lambda}+8N\dot{\lambda}^{2}+N\dot{\beta}_{1}^{2}+N\dot{\beta}_{2}%
^{2}\right) \label{B4b}%
\end{equation}
and \ $R^{\ast}=R^{\ast}\left(  \lambda,\beta_{1},\beta_{2}\right)  $ is the
component of the three dimensional hypersurface which depends on the
structure constants of the algebra $N_{1-3}$ of the Killing algebra of the
Bianchi algebras \cite{CapBianchi}. The general form of $R^{\ast}$ is%
\begin{equation}
R^{\ast}=-\frac{1}{2}e^{-2\lambda}\left[
\begin{array}
[c]{c}%
N_{1}^{2}e^{4\beta_{1}}+e^{-2\beta_{2}}\left(  N_{2}e^{\sqrt{3}\beta_{2}%
}-N_{3}e^{-\sqrt{3}\beta_{2}}\right)  ^{2}+\\
-2N_{1}e^{\beta_{1}}\left(  N_{2}e^{\sqrt{3}\beta_{2}}-N_{3}e^{-\sqrt{3}%
\beta_{2}}\right)  ^{2}%
\end{array}
\right]  +\frac{1}{2}N_{1}N_{2}N_{3}(1+N_{1}N_{2}N_{3})\label{B6}%
\end{equation}
and the special forms for the class A spacetimes can be found in Table
\ref{Bianchi3dR}.%
\begin{table}[tbp] \centering
\caption{The Ricci scalar of the 3d hypersurfaces of the class A Bianchi
spacetimes.}%
\begin{tabular}
[c]{cc}\hline\hline
\textbf{Model} & $R^{\ast}\left(  \lambda,\beta_{1},\beta_{2}\right)
$\\\hline
Bianchi I & $0$\\
Bianchi$~$II & $-\frac{1}{2}e^{\left(  4\beta_{1}-2\lambda\right)  }$\\
Bianchi VI$_{0}$/VII$_{0}$ & $-\frac{1}{2}e^{-2\lambda}\left(  e^{4\beta_{1}%
}+e^{-2\left(  \beta_{1}-\sqrt{3}\beta_{2}\right)  }\pm2e^{\beta_{1}+\sqrt
{3}\beta_{2}}\right)  $\\
Bianchi VIII & $-\frac{1}{2}e^{-2\lambda}\left(
\begin{array}
[c]{c}%
e^{4\beta_{1}}+e^{-2\beta_{1}}\left(  e^{\sqrt{3}\beta_{2}}+e^{-\sqrt{3}%
\beta_{2}}\right)  ^{2}+\\
-2e^{-\beta_{1}}\left(  e^{\sqrt{3}\beta_{2}}-e^{-\sqrt{3}\beta_{2}}\right)
^{2}%
\end{array}
\right)  $\\
Bianchi IX & $-\frac{1}{2}e^{-2\lambda}\left(
\begin{array}
[c]{c}%
e^{4\beta_{1}}+e^{-2\beta_{1}}\left(  e^{\sqrt{3}\beta_{2}}-e^{-\sqrt{3}%
\beta_{2}}\right)  ^{2}+\\
-2e^{-\beta_{1}}\left(  e^{\sqrt{3}\beta_{2}}+e^{-\sqrt{3}\beta_{2}}\right)
^{2}%
\end{array}
\right)  +1$\\\hline\hline
\end{tabular}
\label{Bianchi3dR}%
\end{table}%

In the case of General Relativity, when the action of the field
equations is the Einstein-Hilbert action (we consider that the spacetime is
empty)%
\begin{equation}
S_{GR}=\int dx^{4}\sqrt{-g}R\label{B7}%
\end{equation}
the field equations for the Bianchi class A spacetimes follow from the
Lagrangian%
\begin{equation}
L\left(  N,\lambda,\beta_{1},\beta_{2},\dot{\lambda},\dot{\beta}_{1}%
,\dot{\beta}_{2}\right)  =N\left(  t\right)  e^{3\lambda}\left(  6\dot
{\lambda}^{2}-\frac{3}{2}\left(  \dot{\beta}_{1}^{2}+\dot{\beta}_{2}%
^{2}\right)  \right)  +\frac{e^{3\lambda}}{N(t)}R^{\ast}\label{B8}%
\end{equation}
and the corresponding field equations are the Euler-Lagrange equations with
respect to the variables $\left\{  N,\lambda,\beta_{1},\beta_{2}\right\}  $.
The Euler-Lagrange equations for the variables $\beta_{1},\beta_{2}$ are:%
\begin{equation}
\ddot{\beta}_{\left(  1,2\right)  }+\frac{\dot{N}}{N}\dot{\beta}_{\left(
1,2\right)  }+3\dot{\lambda}\dot{\beta}_{\left(  1,2\right)  }+\frac{1}%
{3N^{2}}R_{,\left(  1,2\right)  }^{\ast}=0\label{B9}%
\end{equation}
for the variable $\lambda$:%
\begin{equation}
4\ddot{\lambda}+\left(  6\dot{\lambda}^{2}+\frac{3}{2}(\dot{\beta_{1}}%
^{2}+\dot{\beta_{2}}^{2})+\frac{1}{2}\dot{\phi}^{2}\right)  +\frac{\dot{N}}%
{N}\dot{\lambda}-\frac{1}{N^{2}}\left(  R^{\ast}+\frac{1}{3}\frac{\partial
R^{\ast}}{\partial\lambda}\right)  =0\label{B10}%
\end{equation}
and for the variable $N$~we have the $G_{0}^{0}=0$ Einstein equation%
\begin{equation}
N\left(  t\right)  e^{3\lambda}\left(  6\dot{\lambda}^{2}-\frac{3}{2}\left(
\dot{\beta}_{1}^{2}+\dot{\beta}_{2}^{2}\right)  \right)  -\frac{e^{3\lambda}%
}{N(t)}R^{\ast}=0.\label{B11}%
\end{equation}

Lagrangian (\ref{B8}) is  singular since $\frac{\partial
L}{\partial\dot{N}}=0$, however, if we consider that
\[
N\left(  t\right)  =N_{0}\text{ or~}N\left(  t\right)  =N\left(
\lambda\left(  t\right)  ,\beta_{1}\left(  t\right)  ,\beta_{2}\left(
t\right)  \right)
\]
then Lagrangian (\ref{B8}) becomes a regular time independent Lagrangian which
admits always, as a Noether integral, the Hamiltonian constant. Hence, equation
(\ref{B11}) can be seen as the energy constrain of the field equations.

In the following we will quantize equation (\ref{B11}) in order to write the
Wheeler-DeWitt (WDW) equation and to perform a symmetry analysis using the Lie
point symmetries in the case of General Relativity and minimally coupled scalar tensor cosmology.

\section{Symmetries of the WDW equation in General Relativity}

\label{BianchiWDW}

In order to simplify equation (\ref{B11}), we consider $N\left(  t\right)
=\bar{N}\left(  t\right)  e^{-3\lambda}$ in the line element (\ref{B2}).
Furthermore, we consider the following change of the variables $\left(
\lambda,\beta_{1},\beta_{2}\right)  =\left(  \frac{\sqrt{3}}{6}x,\frac
{\sqrt{3}}{3}y,\frac{\sqrt{3}}{3}z\right)  $, then the Lagrangian (\ref{B8})
becomes
\begin{equation}
L=\frac{1}{2}\bar{N}\left(  \dot{x}^{2}-\dot{y}^{2}-\dot{z}^{2}\right)
+\frac{1}{\bar{N}}e^{\sqrt{3}x}R^{\ast}.\label{B12}%
\end{equation}

Therefore, we define the momentum $p_{\left(  x,y,z\right)  }=\frac{\partial
L}{\partial\left(  \dot{x},\dot{y},\dot{z}\right)  }$ and equation (\ref{B11})
has now the form%
\begin{equation}
\frac{1}{2}\left(  p_{x}^{2}-p_{y}^{2}-p_{z}^{2}\right)  -e^{\sqrt{3}x}%
R^{\ast}=0\mathbf{.}\label{B13}%
\end{equation}

Equation (\ref{B13}) can be seen as the Hamiltonian of a particle moving in
the space $M^{3}$ with potential $V\left(  x,y,z\right)  =-e^{\sqrt{3}%
x}R^{\ast}$. Furthermore, the field equations are the Hamiltonian constraint (\ref{B13})
and the Hamilton equations
\begin{equation}
\dot{x}=\frac{1}{\bar{N}}p_{x}~,~\dot{y}=-\frac{1}{\bar{N}}p_{y},~\dot
{z}=-\frac{1}{\bar{N}}p_{z}\label{B13A}%
\end{equation}%
\begin{equation}
\dot{p}_{x}=-\sqrt{3}\left(  1+\left(  \ln R^{\ast}\right)  _{x}\right)
e^{\sqrt{3}x}R^{\ast}\label{B13B}%
\end{equation}%
\begin{equation}
\dot{p}_{y}=e^{\sqrt{3}x}R_{,y}^{\ast}~,~\dot{p}_{z}=e^{\sqrt{3}x}R_{,z}%
^{\ast}\label{B13C}%
\end{equation}
Since the minisuperspace is flat; that is the Ricci scalar vanishes, the WDW
equation can be achieved  by a standard  quantization, assuming the conjugate momenta $p_{i}=\frac{\partial
L}{\partial x^{i}}$. Hence from (\ref{B13}),  we have the WDW equation of the form%
\begin{equation}
\Psi_{,xx}-\Psi_{,yy}-\Psi_{,zz}-2e^{\sqrt{3}x}R^{\ast}\Psi=0.\label{B14}%
\end{equation}
which is nothing else but the  Klein Gordon equation in the $M^{3}$ space.

In order to determine the Lie symmetries of (\ref{B14}),  we will use the
geometric results in Ref. \cite{AnIJGMMP}. The $M^{3}$ spacetime admits a ten
dimensional conformal algebra. In particular, it admits a six dimensional
Killing algebra which is the $T^{3}\cup SO\left(  3\right)  $ with elements:
\[
K_{\left(  x\right)  }=\partial_{x}~,K_{\left(  y\right)  }=\partial
_{y}~,~K_{\left(  z\right)  }=\partial_{z}%
\]%
\[
R_{\left(  xy\right)  }=y\partial_{x}+x\partial_{y}~,~R_{\left(  xz\right)
}=z\partial_{x}+x\partial_{z}~,~R_{\left(  yz\right)  }=z\partial
_{y}-y\partial_{z}%
\]
one gradient homothetic Killing vector (HV)%
\[
H=x\partial_{x}+y\partial_{y}+z\partial_{z}~,~\psi_{H}=1
\]
and three special conformal Killing vectors (CKVs) which are%
\begin{align*}
C_{\left(  x\right)  }  & =\frac{1}{2}\left(  x^{2}+y^{2}+z^{2}\right)
\partial_{x}+xy\partial_{y}+xz\partial_{z}~,~\psi_{\left(  x\right)  }=x\\
C_{\left(  y\right)  }  & =xy\partial_{x}+\frac{1}{2}\left(  x^{2}+y^{2}%
-z^{2}\right)  \partial_{y}+zy\partial_{z}~,~\psi_{\left(  y\right)  }=y\\
C_{\left(  z\right)  }  & =xz\partial_{x}+yz\partial_{y}+\frac{1}{2}\left(
x^{2}+z^{2}-y^{2}\right)  \partial_{z}~,~\psi_{\left(  z\right)  }=z\,.
\end{align*}
See Ref. \cite{AnIJGMMP} for details. 
Furthermore, by applying the results in  \cite{AnIJGMMP},  we find that the WDW equation
(\ref{B14}) admits: $i)$ for the Bianchi I model, eleven Lie symmetries, $ii)$ for the
Bianchi II model,  five Lie symmetries, $iii)$ two Lie symmetries for the models
VI$_{0}$/VII$_{0}$ and $iv)$ one Lie symmetry, the linear one, for the models VIII
and IX. In Table \ref{BianchiGR},  we give the corresponding Lie symmetries of
the WDW equation (\ref{B14}) for each Bianchi model.%

\begin{table}[tbp] \centering
\caption{Lie symmetries of the WDW equation of the Class A Bianchi models in
General Relativity}%
\begin{tabular}
[c]{ccc}\hline\hline
\textbf{Model} & $\#~$ & \textbf{Lie Symmetries}\\\hline
Bianchi I & $11$ & $\Psi\partial_{\Psi},~K_{\left(  x\right)  },~K_{\left(
y\right)  },~K_{\left(  z\right)  },~R_{\left(  xy\right)  },~R_{\left(
xz\right)  },~R_{\left(  yz\right)  },~H$\\
&  & $\left(  C_{\left(  x\right)  }-\frac{1}{2}x\Psi\partial_{\Psi}\right)
,\left(  C_{\left(  y\right)  }-\frac{1}{2}y\Psi\partial_{\Psi}\right)
,~\left(  C_{\left(  z\right)  }-\frac{1}{2}z\Psi\partial_{\Psi}\right)  $\\
Bianchi$~$II & $4$ & $\Psi\partial_{\Psi},~K_{\left(  z\right)  },~K_{\left(
x\right)  }-\frac{1}{2}K_{\left(  y\right)  },~R_{\left(  xz\right)  }%
-\frac{1}{2}R_{\left(  yz\right)  }$\\
Bianchi VI$_{0}$/VII$_{0}$ & $2$ & $\Psi\partial_{\Psi},~K_{\left(  x\right)
}-\frac{1}{2}K_{\left(  y\right)  }+\frac{\sqrt{3}}{2}K_{\left(  z\right)  }%
$\\
Bianchi VIII/IX & $1$ & $\Psi\partial_{\Psi}$\\\hline\hline
\end{tabular}
\label{BianchiGR}%
\end{table}%

\section{Invariant solutions of the WDW equation in General Relativity}

\label{BianchiWDWI}

In this section,  we will apply the zero$^{th}$ order invariants of the Lie point
symmetries to reduce the order of the WDW equation (\ref{B14}) and, if
possible,  to determine invariant solutions.
From the results of Table \ref{BianchiGR},  we can see that it is possible to
find invariant solutions for the WDW equation for  Bianchi I and II spacetimes.

\subsection{Bianchi I cosmology}

Since for  Bianchi I spacetimes hold the property  that $R^{\ast}=0$, the WDW equation
(\ref{B14}) is the (1+2) wave equation. The reduction of the (1+2) wave equation
and the invariant solutions have been studied in \cite{Abraham} and recently
in \cite{Type2Laplace}. However, we want to give a concrete  example of 
application of the Lie symmetries.

Let us  consider the Lie algebra $\left\{  X_{\left(  x\right)  }^{I},X_{\left(
y\right)  }^{I}\right\} $ with commutators $\left[  X_{\left(  x\right)  }%
^{I},X_{\left(  y\right)  }^{I}\right]  =0$ where%
\begin{equation}
X_{\left(  x^{i}\right)  }^{I}=K_{\left(  x^{i}\right)  }+\mu_{i}\Psi
\partial_{\Psi}%
\end{equation}
hence the solution of equation~(\ref{B14}) is
\begin{equation}
\Psi_{I}\left(  x,y,z\right)  =\exp\left( \mu_{1}x+\mu_{2}y\right)  \left[
c_{1}\exp\left(  \sqrt{\mu_{1}^{2}-\mu_{2}^{2}}z\right)  +c_{2}\exp\left(
-\sqrt{\mu_{1}^{2}-\mu_{2}^{2}}z\right)  \right]  .
\end{equation}
One can also consider the Lie algebra $\left\{  X_{\left(  x\right)  }%
^{I},X_{\left(  yz\right)  }^{I}=R_{\left(  yz\right)  }+\mu\Psi\partial
_{\Psi}\right\}  $ for which the invariant solution is%
\begin{equation}
\Psi_{I}\left(  x,y,z\right)  =\left[  c_{1}I_{i\mu}\left(  \mu_{1}\sqrt
{y^{2}+z^{2}}\right)  +c_{2}K_{i\mu}\left(  \left(  \mu_{1}\sqrt{y^{2}+z^{2}%
}\right)  \right)  \right]  \exp\left(  \mu\arctan\frac{y}{z}+\mu_{1}x\right)
\end{equation}
where $I_{i\mu},K_{i\mu}$ are the modified Bessel functions of the first and
the second kinds. Similarly,  for the others Lie sub-algebras,  we can construct
invariant solutions. We want also to  recall that the WDW equation
is a linear second order partial differential equation and any linear
combination of the solutions is also a solution.

\subsubsection{The WKB approximation and the classical solution}

Adopting  the WKB approximation to the WDW equation,  equation (\ref{B14})
becomes%
\begin{equation}
\left(  \frac{\partial S}{\partial x}\right)  ^{2}-\left(  \frac{\partial
S}{\partial y}\right)  ^{2}-\left(  \frac{\partial S}{\partial z}\right)
^{2}=0\label{B1.1}%
\end{equation}
which is the null Hamilton-Jacobi equation of the Hamiltonian system
(\ref{B13})-(\ref{B13C}), where $p_{i}=\frac{\partial S}{\partial x^{i}}$. The
solution of (\ref{B1.1}) is%
\begin{equation}
S\left(  x,y,z\right)  =s_{1}y+s_{2}z+\varepsilon\sqrt{s_{1}^{2}+s_{2}^{2}%
}~x+s_{0}, \quad \mbox{with } \quad \varepsilon=\pm1\mathbf{.}%
\end{equation}

Hence the Hamilton equations (\ref{B13A})-(\ref{B13C}) reduce to the
following system (where $\bar{N}=1$)%
\begin{equation}
\dot{x}=\varepsilon\sqrt{s_{1}^{2}+s_{2}^{2}}~,~\dot{y}=-s_{1}~,~\dot
{z}=-s_{2}%
\end{equation}
and therefore we have the solutions
\begin{equation}
\left(  \lambda\left(  t\right)  ,\beta_{1}\left(  t\right)  ,\beta_{2}\left(
t\right)  \right)  =\left(  \frac{\sqrt{3}}{6}\varepsilon\sqrt{s_{1}^{2}%
+s_{2}^{2}}t,-\frac{\sqrt{3}}{3}s_{1}t,-\frac{\sqrt{3}}{3}s_{2}t\right)
\end{equation}
and, by  the coordinate transformation $d\tau=e^{3\lambda}dt$ in the line
element (\ref{B2}),  we obtain  a Kasner spacetime.

\subsection{Bianchi II cosmology}

For the Bianchi II spacetime,  using Table \ref{BianchiGR},  we have that if
we use the Lie symmetries $\left\{  X_{\left(  z\right)  }^{I},~X_{\left(
x\right)  }^{I}-\frac{1}{2}X_{\left(  y\right)  }^{I}\right\}  $ with zero
commutator, the invariant solution is%
\begin{equation}
\Psi_{II}^{1}\left(  x,y,z\right)  =\exp\left(  \frac{2\mu_{12}}{3}\left(
y+2x\right)  +\mu_{3}z\right)  \left(  c_{1}I_{c_{0}}\left(  u\left(
x,y\right)  \right)  +c_{2}K_{c_{0}}\left(  u\left(  x,y\right)  \right)
\right)
\end{equation}
where $\mu_{12}=\mu_{1}-\mu_{2}$,~$c_{0}=\frac{\sqrt{12\mu_{12}^{2}-9\mu
_{3}^{2}}}{3}$, $u\left(  x,y\right)  =\exp\left(  \frac{\sqrt{3}\left(
2y+x\right)  }{3}\right)  $ and $I_{c_{0}},K_{c_{0}}$ are the modified Bessel
functions of the first and the second kinds, respectively. For the Lie
algebra
\[
K_{\left(  x\right)  }^{I}-\frac{1}{2}K_{\left(  y\right)  }^{I},R_{\left(
xz\right)  }-\frac{1}{2}R_{\left(  yz\right)  }%
\]
we have the invariant solution%
\begin{equation}
\Psi_{II}^{2}\left(  x,y,z\right)  =\left(  c_{1}I_{0}\left(  u\left(
x,y\right)  \right)  +c_{2}K_{0}\left(  u\left(  x,y\right)  \right)  \right)
\mathbf{.}%
\end{equation}
The last solution $\Psi_{II}^{2}$ is included in solution $\Psi_{II}^{1}$ for
$\mu_{12}=\mu_{3}=0$.

\subsubsection{The WKB approximation and the classical solution}

\label{bianchiIIwkb}

One can also apply the WKB approximation for the Bianchi II model to the
equation (\ref{B14}):%
\begin{equation}
\left(  \frac{\partial S}{\partial x}\right)  ^{2}-\left(  \frac{\partial
S}{\partial y}\right)  ^{2}-\left(  \frac{\partial S}{\partial z}\right)
^{2}+e^{\frac{2\sqrt{3}}{3}\left(  2y+x\right)  }=0\label{nHJB2}%
\end{equation}
Let us adopt, the coordinate transformation $y=w-\frac{x}{2}$ in the Hamiltonian
system. Hence, the new Hamilton-Jacobi equation (\ref{nHJB2}) becomes%
\begin{equation}
\left(  \frac{\partial S}{\partial x}\right)  ^{2}+\left(  \frac{\partial
S}{\partial x}\right)  \left(  \frac{\partial S}{\partial w}\right)  -\frac
{3}{4}\left(  \frac{\partial S}{\partial w}\right)  ^{2}-\left(
\frac{\partial S}{\partial z}\right)  ^{2}+e^{\frac{4\sqrt{3}}{3}w}=0
\end{equation}
with the solution%
\begin{equation}
S\left(  x,w,z\right)  =S_{1}\left(  x\right)  +S_{2}\left(  w\right)
+S_{3}\left(  z\right)  \mathbf{,}%
\end{equation}
where the functions $S_{1,2,3}$ follow from the  system
\begin{equation}
S_{1}\left(  x\right)  =c_{1}x~,~S_{3}\left(  z\right)  =c_{2}z
\end{equation}%
\begin{equation}
\left(  \frac{dS_{2}}{dw}\right)  ^{2}=\frac{4}{3}\left[  c_{1}\left(
c_{1}+\frac{dS_{2}}{dw}\right)  +c_{2}^{2}-2e^{\frac{4\sqrt{3}}{3}w}\right]
=0\mathbf{.}%
\end{equation}
The Hamilton function $S\left(  x,w,z\right)  $ is
\begin{equation}
S\left(  x,w,z\right)  =c_{1}x+c_{2}z+\frac{2}{3}c_{1}w-\varepsilon\frac
{\sqrt{3}}{3}\left(  M\left(  w\right)  -\sqrt{c_{12}}\arctan\left(
\frac{M\left(  w\right)  }{\sqrt{c_{12}}}\right)  \right) , \quad \mbox{with}\quad \varepsilon=\pm1
\end{equation}
where $c_{12}=3c_{2}^{2}-4c_{1}^{2}$ and$~M\left(  w\right)  =\sqrt
{3e^{\frac{4\sqrt{3}}{3}w}-c_{12}}$. Therefore the reduced Hamiltonian system
is (recall that $\bar{N}=1$)%
\begin{equation}
\dot{z}=-2c_{2}~,~\dot{w}=\varepsilon M\left(  w\right) \label{nHJB3}%
\end{equation}%
\begin{equation}
\dot{x}=\frac{8}{3}c_{1}-\varepsilon\frac{2}{3}M\left(  w\right) \label{nHJB4}%
\end{equation}
Hence the solution of the system (\ref{nHJB3}),(\ref{nHJB4}) is%
\begin{equation}
z\left(  t\right)  =-2c_{2}t
\end{equation}%
\begin{equation}
w\left(  t\right)  =-\frac{\sqrt{3}}{4}\ln\left(  \frac{\cos^{2}\left(
\frac{2\sqrt{3}}{3}\varepsilon\sqrt{c_{12}}\left(  t+w_{0}\right)  \right)
}{c_{12}}\right)
\end{equation}%
\begin{equation}
x\left(  t\right)  =\frac{\sqrt{3}}{6}\ln\left[  \cos^{2}\left(  \frac
{2\sqrt{3}}{3}\varepsilon\sqrt{c_{12}}\left(  t+w_{0}\right)  \right)
\right] +\frac{8}{3}c_{1}t+x_{0}\,.
\end{equation}
which is the solution of the empty Bianchi II spacetime in General Relativity.

In the following section we will apply the same procedure in the case of minimally coupled  
scalar-tensor  cosmology. Furthermore, we will use the Lie symmetries of the WDW equation in
order to determine the unknown potential of the scalar field.

\section{Symmetries of the WDW equation in minimally coupled scalar-tensor  cosmology}

\label{BianchiSF}

Let us continue  the Lie symmetry analysis of the WDW equation for 
cosmological containing a minimally coupled  scalar field in the gravitational action. The Noether symmetry classification of the field equations has been
studied in \cite{CapB,Cotsakis,VakB} . Noether symmetries  have been adopted in the jet space in \cite{GRG}.  In Ref.\cite{kamen}, a detailed study of integrable cosmological models with 
non-minimally coupled scalar fields is presented. 

Let us now take into account   the following action:%
\begin{equation}
S=\int dx^{4}\sqrt{-g}\left(  R-\frac{1}{2}g^{\mu\nu}\phi_{,\mu}\phi_{\nu
}+V\left(  \phi\right)  \right)\,  .\label{B15}%
\end{equation}
From the line element (\ref{B2}) and equations (\ref{B4a}),
(\ref{B4b}) we find the Lagrangian  \cite{DemB}
\begin{equation}
L\left(  N,\lambda,\beta_{1},\beta_{2},\phi,\dot{\lambda},\dot{\beta}_{1}%
,\dot{\beta}_{2},\dot{\phi}\right)  =N(t)e^{3\lambda}\left(  6\dot{\lambda
}^{2}-\frac{3}{2}\left(  \dot{\beta}_{1}^{2}+\dot{\beta}_{2}^{2}\right)
-\frac{1}{2}\dot{\phi}^{2}\right)  +\frac{e^{3\lambda}}{N(t)}\left(  V\left(
\phi\right)  +R^{\ast}\right)  .\label{B16}%
\end{equation}
Hence, by applying the Euler-Lagrange vector in (\ref{B16}),  we find four field
equations which are the two equations (\ref{B9}) and
\begin{equation}
4\ddot{\lambda}+\left(  6\dot{\lambda}^{2}+\frac{3}{2}(\dot{\beta_{1}}%
^{2}+\dot{\beta_{2}}^{2})+\frac{1}{2}\dot{\phi}^{2}\right)  +\frac{\dot{N}}%
{N}\dot{\lambda}-\frac{1}{N^{2}}\left(  V+R^{\ast}+\frac{1}{3}\frac{\partial
R^{\ast}}{\partial\lambda}\right)  =0\label{B17}%
\end{equation}

\begin{equation}
Ne^{3\lambda}\left(  6\dot{\lambda}^{2}-\frac{3}{2}\left(  \dot{\beta}_{1}%
^{2}+\dot{\beta}_{2}^{2}\right)  -\frac{1}{2}\dot{\phi}^{2}\right)
-\frac{e^{3\lambda}}{N}\left(  V+R^{\ast}\right)  =0.\label{B18}%
\end{equation}
The last  corresponds to the  the $G_{0}^{0}=T_{0}^{0}$ Einstein equation. Furthermore, from the
Euler-Lagrange equation ${\displaystyle \frac{d}{dt}\left(  \frac{\partial L}{\partial
\dot{\phi}}\right)  -\frac{\partial L}{\partial\phi}}$, we obtain the field
equation for the scalar field%
\begin{equation}
\ddot{\phi}+3\dot{\lambda}\dot{\phi}+\frac{\dot{N}}{N}\dot{\phi}+\frac
{1}{N^{2}}\frac{\partial V}{\partial\phi}=0.\label{B19}%
\end{equation}
which can also be derived by  the Bianchi identity $T_{~~;\nu}^{\mu\nu}=0$,
where $T^{\mu\nu}$ is the energy-momentum tensor for scalar field.
As in the case of General Relativity,   the coordinate transformations
$\left(  \lambda,\beta_{1},\beta_{2}\right)  =\left(  \frac{\sqrt{3}}%
{6}x,\frac{\sqrt{3}}{3}y,\frac{\sqrt{3}}{3}z\right)  $ and $N\left(  t\right)
=\bar{N}\left(  t\right)  e^{-3\lambda},$ can be adopted.  Hence, equation (\ref{B18}) becomes%
\begin{equation}
\frac{1}{2}\bar{N}\left(  \dot{x}^{2}-\dot{y}^{2}-\dot{z}^{2}-\dot{\phi}%
^{2}\right)  -\frac{1}{\bar{N}}e^{\sqrt{3}x}\left(  V\left(  \phi\right)
+R^{\ast}\right)  =0\,.\label{B20}%
\end{equation}
Moreover, by using the momentum ${\displaystyle p_{\left(  x,y,z,\phi\right)  }%
=\frac{\partial L}{\partial x^{i}}}$,  equation (\ref{B20}) becomes%
\begin{equation}
\frac{1}{2}\left(  p_{x}^{2}-p_{y}^{2}-p_{z}^{2}-p_{\phi}^{2}\right)
-e^{\sqrt{3}x}\left(  V\left(  \phi\right)  +R^{\ast}\right)  =0\label{B21}%
\end{equation}
and hence, from (\ref{B21}), we have the following WDW equation%
\begin{equation}
\Psi_{,xx}-\Psi_{,yy}-\Psi_{,zz}-\Psi_{,\phi\phi}-2e^{\sqrt{3}x}\left(
V\left(  \phi\right)  +R^{\ast}\right)  \Psi=0\label{B22}%
\end{equation}
since the minisuperspace of equation (\ref{B21}) is the flat space $M^{4}$. We
note that equation (\ref{B22}) is the Klein Gordon equation in  $M^{4}$.

Since we want to use the geometric approach in Ref. \cite{AnIJGMMP},  we need the
conformal algebra of the $M^{4}$ spacetime. The four dimensional flat space
$M^{4}$ admits ten Killing vectors (KVs) which are%

\[
K_{\left(  x\right)  },~K_{\left(  y\right)  },~K_{\left(  z\right)
},~K_{\left(  \phi\right)  }=\partial_{\phi}\mathbf{,}%
\]%
\[
R_{\left(  xy\right)  },~R_{\left(  xz\right)  },~R_{\left(  yz\right)
},~R_{\left(  x\phi\right)  }=\phi\partial_{x}+x\partial_{\phi}\mathbf{,}%
\]%
\[
R_{\left(  y\phi\right)  }=\phi\partial_{y}-y\partial_{\phi}~,~R_{\left(
z\phi\right)  }=\phi\partial_{z}-z\partial_{\phi}\mathbf{,}%
\]
one gradient HV%
\[
\bar{H}=x\partial_{x}+y\partial_{y}+z\partial_{z}+\phi\partial_{\phi
}\mathbf{,}%
\]
and four special CKVs%

\begin{align*}
\bar{C}_{\left(  x\right)  }  & =C_{\left(  x\right)  }+\frac{1}{2}\phi
^{2}\partial_{x}+\phi x\partial_{\phi}~,~\psi_{\left(  x\right)  }=x\\
\bar{C}_{\left(  y\right)  }  & =C_{\left(  y\right)  }-\frac{1}{2}\phi
^{2}\partial_{y}+\phi y\partial_{\phi}~,~\psi_{\left(  y\right)  }=y\\
\bar{C}_{\left(  z\right)  }  & =C_{\left(  z\right)  }-\frac{1}{2}\phi
^{2}\partial_{z}+\phi z\partial_{\phi}~,~\psi_{\left(  z\right)  }=z
\end{align*}%
\[
\bar{C}_{\left(  \phi\right)  }=x\phi\partial_{x}+y\phi\partial_{y}%
+z\phi\partial_{z}+\frac{1}{2}\left(  x^{2}+\phi^{2}-y^{2}-z^{2}\right)
\partial_{\phi}~,~\psi_{\left(  \phi\right)  }=\phi
\]
One can find that the WDW equation (\ref{B22}) admits Lie symmetries for an
arbitrary potential $V\left(  \phi\right)  $; in the case of Bianchi I, 
spacetime equation (\ref{B22}) admits four Lie symmetries; for Bianchi II, one
has two Lie symmetries and one Lie symmetry for the rest of the Bianchi
models. Therefore, for special forms of potentials $V\left(  \phi\right) $, it
is possible that the WDW equation (\ref{B22}) admits extra symmetries. The
special forms of the potentials that we found are: (a) $V\left(  \phi\right)
=0$, in which the scalar field behaves as stiff matter; (b) $V\left(  \phi\right)
=V_{0}$ with $V_{0}\neq0$ and (c) $V\left(  \phi\right)  =V_{0}e^{\mu\phi}$.
The results of symmetry analysis are collected in Tables \ref{BianchiSFV0} and
\ref{BianchiSFVphi}.%

\begin{table}[tbp] \centering
\caption{Lie symmetries of the WDW equation of the Class A Bianchi models in
scalar field cosmology for  $V(\phi)=0$.}%
\begin{tabular}
[c]{ccc}\hline\hline
\textbf{Model~}$V\left(  \phi\right)  =0$ & $\#~$ & \textbf{Lie Symmetries}%
\\\hline
Bianchi I & $16$ & $\Psi\partial_{\Psi},~K_{\left(  x\right)  },~K_{\left(
y\right)  },~K_{\left(  z\right)  },~K_{\left(  \phi\right)  },~R_{\left(
xy\right)  },~R_{\left(  xz\right)  }$\\
&  & $~R_{\left(  yz\right)  },~R_{\left(  x\phi\right)  },~R_{\left(
y\phi\right)  },~R_{\left(  z\phi\right)  },~\bar{H},~\left(  \bar{C}_{\left(
x\right)  }-x\Psi\partial_{\Psi}\right)  ,~$\\
&  & $\left(  \bar{C}_{\left(  y\right)  }-y\Psi\partial_{\Psi}\right)
,~\left(  \bar{C}_{\left(  z\right)  }-z\Psi\partial_{\Psi}\right)  ,~\left(
\bar{C}_{\left(  \phi\right)  }-\phi\Psi\partial_{\Psi}\right)  $\\
Bianchi$~$II & $7$ & $\Psi\partial_{\Psi},~K_{\left(  z\right)  },~K_{\left(
\phi\right)  },~K_{\left(  x\right)  }-\frac{1}{2}K_{\left(  y\right)
},~R_{\left(  z\phi\right)  }$\\
&  & $R_{\left(  xz\right)  }-\frac{1}{2}R_{\left(  yz\right)  },~R_{\left(
x\phi\right)  }-\frac{1}{2}R_{\left(  y\phi\right)  }$\\
Bianchi VI$_{0}$/VII$_{0}$ & $3$ & $\Psi\partial_{\Psi},~K_{\left(
\phi\right)  },~K_{\left(  x\right)  }+\frac{1}{4}K_{\left(  y\right)  }%
+\frac{\sqrt{3}}{4}K_{\left(  z\right)  }$\\
&  & $R_{\left(  x\phi\right)  }+\frac{1}{4}R_{\left(  y\phi\right)  }%
+\frac{\sqrt{3}}{4}R_{\left(  z\phi\right)  }$\\
Bianchi VIII/IX & $2$ & $\Psi\partial_{\Psi},~K_{\left(  \phi\right)  }%
$\\\hline\hline
\end{tabular}
\label{BianchiSFV0}%
\end{table}%
%

\begin{table}[tbp] \centering
\caption{Lie symmetries of the WDW equation of the Class A Bianchi models in
scalar field cosmology for non-zero potentials.}%
\begin{tabular}
[c]{ccc}\hline\hline
\textbf{Model}~$V\left(  \phi\right)  =V_{0}$ & $\#~$ & \textbf{Lie
Symmetries}\\\hline
Bianchi I & $7$ & $\Psi\partial_{\Psi},~K_{\left(  y\right)  },~K_{\left(
z\right)  },~K_{\left(  \phi\right)  },~R_{\left(  yz\right)  },~R_{\left(
y\phi\right)  },~R_{\left(  z\phi\right)  }~$\\
Bianchi$~$II & $4$ & $\Psi\partial_{\Psi},~K_{\left(  z\right)  },~K_{\left(
\phi\right)  },~R_{\left(  z\phi\right)  }~$\\
Bianchi VI$_{0}$/VII$_{0}$ & $2$ & $\Psi\partial_{\Psi},~K_{\left(
\phi\right)  }$\\
Bianchi VIII/IX & $2$ & $\Psi\partial_{\Psi},~K_{\left(  \phi\right)  }$\\
&  & \\
\textbf{Model}~$V\left(  \phi\right)  =V_{0}e^{\mu\phi}$ & $\#~$ & \textbf{Lie
Symmetries}\\
Bianchi I & $7$ & $\Psi\partial_{\Psi},K_{\left(  y\right)  },~K_{\left(
z\right)  },~R_{\left(  yz\right)  },~\frac{\sqrt{3}}{3}\mu K_{(x)}-K_{(\phi
)},~~$\\
&  & $R_{\left(  y\phi\right)  }+\frac{\sqrt{3}}{3}\mu R_{\left(  xy\right)
},~R_{\left(  z\phi\right)  }+\frac{\sqrt{3}}{3}\mu R_{\left(  xz\right)  }$\\
Bianchi$~$II & $4$ & $\Psi\partial_{\Psi},~K_{\left(  z\right)  },~K_{\left(
x\right)  }-\frac{1}{2}K_{\left(  y\right)  }-\frac{\sqrt{3}}{\mu}K_{\left(
\phi\right)  }$\\
&  & $R_{\left(  z\phi\right)  }+\frac{\sqrt{3}}{3}\mu\left(  R_{\left(
xz\right)  }-\frac{1}{2}R_{\left(  yz\right)  }\right)  $\\
Bianchi VI$_{0}$/VII$_{0}$ & $2$ & $\Psi\partial_{\Psi},~K_{\left(  x\right)
}-\frac{1}{2}K_{\left(  y\right)  }-\frac{\sqrt{3}}{2}K_{\left(  z\right)
}-\frac{\sqrt{3}}{\mu}K_{\left(  \phi\right)  }$\\
Bianchi VIII/IX & $1$ & $\Psi\partial_{\Psi}$\\
&  & \\
\textbf{Model}~$V\left(  \phi\right)  =V\left(  \phi\right)  $ & $\#~$ &
\textbf{Lie Symmetries}\\
Bianchi I & $4$ & $\Psi\partial_{\Psi},~K_{\left(  y\right)  },~K_{\left(
z\right)  },~R_{\left(  yz\right)  }~$\\
Bianchi$~$II & $2$ & $\Psi\partial_{\Psi},~K_{\left(  z\right)  }~$\\
Bianchi VI$_{0}$/VII$_{0}$ & $1$ & $\Psi\partial_{\Psi}$\\
Bianchi VIII/IX & $1$ & $\Psi\partial_{\Psi}$\\\hline\hline
\end{tabular}
\label{BianchiSFVphi}%
\end{table}%

We will continue our analysis using the results of Tables \ref{BianchiSFV0}
and \ref{BianchiSFVphi} in order to determine invariant solutions of the WDW
equation (\ref{B22}) for cases where it is possible.

\section{Invariant solutions of the WDW equation in scalar field cosmology}

\label{BianchiSFWDW}

From the symmetries in Tables \ref{BianchiSFV0} and \ref{BianchiSFVphi}, we
observe that invariant solutions for the WDW equation can be determined for the
Bianchi type I model for (a) $V\left(  \phi\right)  =0$, (b) $V\left(\phi\right)  =V_{0},$ (c) $V\left(  \phi\right)  =V_{0}e^{\mu\phi}$; and, for
the Bianchi II model,  for zero potential. It is worth noticing  that, in the
following,  we will choose  $\bar{N}\left(  t\right)  =1$.

\subsection{Bianchi I cosmology}
For the Bianchi I spacetime, the WDW equation (\ref{B22}) becomes%
\begin{equation}
\Psi_{,xx}-\Psi_{,yy}-\Psi_{,zz}-\Psi_{,\phi\phi}-2e^{\sqrt{3}x}V\left(
\phi\right)  \Psi=0\,.\label{BSF.01}%
\end{equation}
If the potential $V\left(  \phi\right)  =0$, then (\ref{BSF.01}) becomes the
(1+3) wave equation in $E^{3}$ \cite{Abraham}, hence we will omit this case.
When $V\left(  \phi\right)  =V_{0},~V_{0}\neq0$, the field equations are
equivalent to the case of General Relativity with stiff matter and a
cosmological constant. In this case,  we  use zero order invariants of the
Lie symmetries
\begin{equation}
\bar{X}_{\left(  A\right)  }=K_{\left(  A\right)  }+\mu_{\left(  A\right)
}\Psi\partial_{\Psi}~,~A=y,z,\phi\label{BSF.02}%
\end{equation}
which form a closed Lie algebra. In this case, equation (\ref{BSF.01})
reduces to the linear second order ordinary differential equation%
\begin{equation}
\Phi^{\prime\prime}-\left(  \mu_{\left(  y\right)  }+\mu_{\left(  z\right)
}+\mu_{\left(  z\right)  }+2V_{0}e^{\sqrt{3}x}\right)  \Phi=0\label{BSF.03}%
\end{equation}
where $\Psi=\Phi\left(  x\right)  \exp\left(  \mu_{\left(  y\right)  }%
y+\mu_{\left(  z\right)  }z+\mu_{\left(  \phi\right)  }\phi\right)  $ and
$\Phi^{\prime}=\frac{d\Phi\left(  x\right)  }{dx}$. Therefore, the solution of
 equation (\ref{BSF.03}) is
\begin{equation}
\Phi\left(  x\right)  =\Phi_{1}J_{c}\left(  i\frac{2\sqrt{6V_{0}}}{3}%
e^{\frac{\sqrt{3}}{2}x}\right)  +\Phi_{2}Y_{c}\left(  i\frac{2\sqrt{6V_{0}}%
}{3}e^{\frac{\sqrt{3}}{2}x}\right) \label{BSF.04}%
\end{equation}
where $J_{c},Y_{c}$ are the Bessel functions of the first and second kind and
the constant is $c=\frac{2\sqrt{3}}{3}\left(  \sqrt{\mu_{\left(  y\right)  }%
^{2}+\mu_{\left(  z\right)  }^{2}+\mu_{\left(  z\right)  }^{2}}\right)  $.
For the exponential potential,  we apply the zero order invariants of the Lie
symmetries
\begin{equation}
\bar{X}_{\left(  y\right)  },\bar{X}_{\left(  z\right)  },~\frac{\sqrt{3}}%
{3}\mu K_{(x)}-K_{(\phi)}+\mathbf{\nu}\Psi\partial_{\Psi}%
\end{equation}
and the WDW equation (\ref{BSF.01}) becomes%
\begin{equation}
\left(  3-\mu^{2}\right)  \Phi^{\prime\prime}\left(  w\right)  +6\nu
\Phi^{\prime}-\left(  \left(  \mu_{\left(  y\right)  }^{2}+\mu_{\left(
z\right)  }^{2}\right)  \mu^{2}-3\mathbf{\nu}^{2}+2V_{0}\mu^{2}e^{\mu
w}\right)  \Phi=0\label{BSF.05}%
\end{equation}
where $\Psi\left(  x,y,z,\phi\right)  =\Phi\left(  w\right)  \exp\left(
\frac{\sqrt{3}\mathbf{\nu}}{\mu}x+\mu_{\left(  y\right)  }y+\mu_{\left(
z\right)  }z\right)  $, $\Phi^{\prime}=\frac{d\Phi\left(  w\right)  }{dw}$ and
$w=\frac{\sqrt{3}}{\mu}x+\phi$.
Hence, for various values of the constant $\mu$ from (\ref{BSF.05}),  we have that%

\begin{equation}
\Phi\left(  w\right)  =\exp\left(\frac{3\mu w}{\mu^{2}-3}\right)\left[  \Phi_{1}J_{\bar{c}%
}\left(  2\sqrt{\frac{2V_{0}}{\mu^{2}-3}}e^{\frac{\mu}{2}w}\right)  +\Phi
_{2}Y_{\bar{c}}\left(  2\sqrt{\frac{2V_{0}}{\mu^{2}-3}}e^{\frac{\mu}{2}%
w}\right)  \right] ,\quad \text{with }\quad \mu\neq\mathbf{\sqrt{3}}\label{BSF.06}%
\end{equation}
where $\bar{c}=\frac{2}{\left\vert \mu^{2}-3\right\vert }\sqrt{3\mathbf{\nu
}^{2}-\left(  \mu^{2}-3\right)  \left(  \mu_{\left(  y\right)  }^{2}%
+\mu_{\left(  z\right)  }^{2}\right)  }$ and
\begin{equation}
\Phi\left(  w\right)  =\Phi_{0}\exp\left[  \frac{1}{2\mathbf{\nu}}\left(
\mu_{\left(  y\right)  }^{2}+\mu_{\left(  z\right)  }^{2}\right)
-\frac{\mathbf{\nu}}{2}w+\frac{\sqrt{3}}{3}\frac{V_{0}}{\mathbf{\nu}}%
e^{\sqrt{3}w}\right]  ~,\text{\ for }\left\vert \mu\right\vert =\sqrt{3}\text{
,~}\mathbf{\nu}\neq0.\label{BSF.07}%
\end{equation}

\subsubsection{The WKB approximation and the classical solutions}

In the WKB approximation the WDW equation (\ref{BSF.01}) becomes the 
Hamilton-Jacobi equation
\begin{equation}
\frac{1}{2}\left[  \left(  \frac{\partial S}{\partial x}\right)  ^{2}-\left(
\frac{\partial S}{\partial y}\right)  ^{2}-\left(  \frac{\partial S}{\partial
z}\right)  ^{2}-\left(  \frac{\partial S}{\partial\phi}\right)  ^{2}\right]
-e^{\sqrt{3}x}V\left(  \phi\right)  =0\label{BSF.08}%
\end{equation}
where $S=S\left(  x,y,z,\phi\right)  ~$describes a motion of a particle in the
$M^{4}$~space. The solution of the  Hamilton Jacobi function leads us to
the following reduced Hamiltonian system%
\begin{equation}
\dot{x}=\frac{\partial S}{\partial x}~,~\dot{y}=\frac{\partial S}{\partial
y}~,~\dot{z}=\frac{\partial S}{\partial z}~,~\dot{\phi}=\frac{\partial
S}{\partial\phi}.\label{BSF.09}%
\end{equation}
For  $V\left(  \phi\right)  =0$, from equation
(\ref{BSF.08}) we have that%
\begin{equation}
S_{0}\left(  x,y,z,\phi\right)  =c_{1}y+c_{2}z+c_{3}\phi+\varepsilon
\sqrt{c_{1}^{2}+c_{2}^{2}+c_{3}^{2}}x~,~\varepsilon=\pm1\text{.}\label{BSF.10}%
\end{equation}
Then from (\ref{BSF.09}) and (\ref{BSF.10}) we have%
\begin{equation}
x\left(  \tau\right)  =\varepsilon\sqrt{c_{1}^{2}+c_{2}^{2}+c_{3}^{2}}t+x_{0}.
\end{equation}%
\begin{equation}
y\left(  t\right)  =-c_{1}t+y_{0},~z\left(  t\right)  =-c_{2}t+z_{0}%
~,~\phi\left(  t\right)  =-c_{3}t+\phi_{0}.
\end{equation}
Similarly, for constant potential, i.e. $V\left(  \phi\right)  =V_{0}$, we
have
\begin{equation}
S_{V_{0}}\left(  x,y,z,\phi\right)  =c_{1}y+c_{2}z+c_{3}\phi+\varepsilon
\frac{2\sqrt{3}}{3}\left(  L\left(  x\right)  +\sqrt{c_{1-3}}\arctan
h\frac{L\left(  x\right)  }{\sqrt{c_{1-3}}}\right)
\end{equation}
where $L\left(  x\right)  =\sqrt{c_{1}^{2}+c_{2}^{2}+c_{3}^{2}+2V_{0}%
e^{\sqrt{3}x}}~$and $c_{1-3}=c_{1}^{2}+c_{2}^{2}+c_{3}^{2}$. Hence the reduced
Hamilton equations (\ref{BSF.09}) are
\begin{equation}
\dot{x}=L\left(  x\right)  ~,~\dot{y}=-c_{1}~,~\dot{z}=-c_{2}~,~\phi=-c_{3}%
\end{equation}
and the exact solution of the field equations is
\begin{equation}
x\left(  t\right)  =\frac{1}{3}\ln\left[  \frac{c_{1-3}}{2V_{0}}\left(
\tanh\left(  \frac{\sqrt{3}}{2}\sqrt{c_{1-3}}\left(  t+x_{0}\right)  \right)
-1\right)  \right]
\end{equation}%
\begin{equation}
y\left(  t\right)  =c_{1}t+y_{0},~z\left(  t\right)  =c_{2}t+z_{0}%
~,~\phi\left(  t\right)  =c_{3}t+\phi_{0}.
\end{equation}
For the exponential potential, $V\left(  \phi\right)  =V_{0}e^{\mu\phi},$ as
we saw previously, there exist different solutions of the WDW equation for
different values of the constant $\mu$. Hence, the solution of the 
Hamilton-Jacobi equation (\ref{BSF.08}) is determined by the various values of
$\mu$.

Let us set $\mu=-\sqrt{3}$. Applying the coordinate transformation $\phi
=\psi+x$ in the Hamiltonian system, the new Hamilton-Jacobi equation is%
\begin{equation}
\frac{1}{2}\left[ \left(  \frac{\partial S}{\partial x}\right)  ^{2}-2\left(
\frac{\partial S}{\partial x}\right)  \left(  \frac{\partial S}{\partial\psi
}\right)  -\left(  \frac{\partial S}{\partial y}\right)  ^{2}-\left(
\frac{\partial S}{\partial z}\right)  ^{2}\right]  -V_{0}e^{-\sqrt{3}\psi
}=0\label{BSF.11A}%
\end{equation}
and the reduced Hamiltonian system is%
\begin{equation}
\dot{x}=\left(  \frac{\partial S}{\partial x}\right)  -\left(  \frac{\partial
S}{\partial\psi}\right)  ~,~\dot{y}=-\frac{\partial S}{\partial y}~,~\dot
{z}=-\frac{\partial S}{\partial z}~,~\dot{\psi}=-\frac{\partial S}{\partial
x}.
\end{equation}
Therefore from (\ref{BSF.11A}) the Hamilton action is
\begin{equation}
S\left(  x,y,z,\psi\right)  =c_{1}x+c_{2}y+c_{3}z+\frac{\left(  c_{2}%
^{2}+c_{3}^{2}-c_{1}^{2}\right)  }{2c_{1}}\psi-\frac{\sqrt{3}V_{0}}{6c_{1}%
}e^{-\sqrt{3}\psi}%
\end{equation}
and the field equations reduce to the system%
\begin{equation}
\dot{x}=\frac{\left(  c_{1}^{2}-c_{2}^{2}-c_{3}^{2}\right)  -V_{0}e^{\sqrt
{3}\psi}}{2c_{1}}~,~\dot{y}=-c_{2}~,~\dot{z}=-c_{3},~\dot{\psi}=-c_{1}%
\end{equation}
with the solution
\begin{equation}
y\left(  t\right)  =-c_{2}t+y_{0}~,~z\left(  t\right)  =-c_{3}t+z_{0}%
~,~\psi\left(  t\right)  =-c_{1}t+\psi_{0}%
\end{equation}
and%
\begin{equation}
x\left(  t\right)  =\frac{3}{2}c_{1}t-\frac{\left(  c_{2}^{2}+c_{3}%
^{2}\right)  }{2c_{1}}t+\frac{\sqrt{3}V_{0}}{6c_{1}^{2}}e^{-\sqrt{3}\psi_{0}%
}e^{\sqrt{3}c_{1}t}+x_{0}\mathbf{.}%
\end{equation}
Furthermore, for $\left\vert \mu\right\vert \neq\sqrt{3}$, we apply the
coordinate transformation $\phi=\psi-\frac{\sqrt{3}}{\mu}x$,
\begin{equation}
\frac{1}{2}\left[  \left(  \frac{\partial S}{\partial x}\right)  ^{2}%
-\frac{2\sqrt{3}}{\mu}\left(  \frac{\partial S}{\partial x}\right)  \left(
\frac{\partial S}{\partial\psi}\right)  -\left(  1-\frac{3}{\mu^{2}}\right)
\left(  \frac{\partial S}{\partial\psi}\right)  ^{2}-\left(  \frac{\partial
S}{\partial y}\right)  ^{2}-\left(  \frac{\partial S}{\partial z}\right)
^{2}\right]  -V_{0}e^{\mu\psi}=0\label{BSF.11}%
\end{equation}
hence from the Hamilton-Jacobi equation (\ref{BSF.08}) we have
\begin{equation}
S\left(  x,y,z,\psi\right)  =c_{1}x+c_{2}y+c_{3}z+\gamma\left(  \psi\right)
\label{BSF.12}%
\end{equation}
where
\begin{equation}
\left(  \frac{d\gamma}{d\psi}\right)  ^{2}=\frac{\mu^{2}}{\mu^{2}-3}\left(
\frac{2\sqrt{3}}{\mu}\frac{d\gamma}{d\psi}-V_{0}e^{\mu\psi}+\left(  c_{1}%
^{2}-c_{2}^{2}-c_{3}^{2}\right)  \right)  .\label{BSF.13}%
\end{equation}

The reduced Hamiltonian system (\ref{BSF.09}) in the new coordinates becomes%
\begin{equation}
~\dot{y}=-\frac{\partial S}{\partial y}~,~\dot{z}=-\frac{\partial S}{\partial
z}\label{BSF.14}%
\end{equation}%
\begin{equation}
\dot{x}=\left(  \frac{\partial S}{\partial x}\right)  -\frac{\sqrt{3}}{\mu
}\left(  \frac{\partial S}{\partial\psi}\right)  ~,~\dot{\psi}=-\frac{\sqrt
{3}}{\mu}\left(  \frac{\partial S}{\partial x}\right)  -\left(  1-\frac{3}%
{\mu^{2}}\right)  \left(  \frac{\partial S}{\partial\psi}\right)
~\label{BSF.15}%
\end{equation}
and therefore from (\ref{BSF.12}) and (\ref{BSF.13}) the last becomes%
\begin{equation}
\dot{y}=-c_{2}~,~\dot{z}=-c_{3}\label{BSF.16}%
\end{equation}%
\begin{equation}
\dot{x}=c_{1}-\frac{3-\sqrt{3}\sqrt{\left(  3-\mu^{2}\right)  c_{1-3}%
V_{0}e^{\mu\psi}}}{\mu^{2}-3}\label{BSF.17}%
\end{equation}%
\begin{equation}
\dot{y}=-\frac{2\sqrt{3}}{\mu}c_{1}-\frac{1}{\mu}\sqrt{\left(  3-\mu
^{2}\right)  c_{1-3}V_{0}e^{\mu\psi}}\mathbf{.}\label{BSF.18}%
\end{equation}
However, if one wants to write an analytical solution of this system, we have
to perform another transformation that is $dt\rightarrow d\tau$. The exact solution of
the exponential potential in Bianchi I scalar field cosmology was found
in \cite{GRG} so we will omit the derivation it in this work.

\subsection{Bianchi II cosmology}

One can observe from Tables \ref{BianchiSFV0} and \ref{BianchiSFVphi} that, for
 Bianchi type II spacetimes,  we can determine invariant solution of the WDW
equation only for zero potential, i.e. the scalar field is a kination fluid
acting  as stiff matter $p_{\phi}=\rho_{\phi}$. In this  case,  the WDW equation
(\ref{B22}) becomes%
\begin{equation}
\Psi_{,xx}-\Psi_{,yy}-\Psi_{,zz}-\Psi_{,\phi\phi}+e^{\frac{2\sqrt{3}}%
{3}\left(  2y+x\right)  }\Psi=0\mathrm{.}\label{BSF.20}%
\end{equation}
By applying the zero order invariants of the following three dimensional
closed Lie algebra with vanishing commutators,%
\begin{equation}
K_{\left(  x\right)  }-\frac{1}{2}K_{\left(  y\right)  }+\mathbf{\nu}%
\Psi\partial_{\Psi},~K_{\left(  z\right)  }+\mu_{\left(  z\right)  }%
\Psi\partial_{\Psi},~K_{\left(  \phi\right)  }+\mu_{\left(  \phi\right)  }%
\Psi\partial_{\Psi}%
\end{equation}
we find the invariant solution%
\begin{equation}
\Psi_{1}\left(  x,y,z,\phi\right)  =\exp\left(  \frac{2\mathbf{\nu}}{3}\left(
y+2x\right)  +\mu_{\left(  z\right)  }z+\mu_{\left(  \phi\right)  }%
\phi\right)  \left(  \Psi_{1}I_{\lambda}\left(  u\left(  x,y\right)  \right)
+\Psi_{2}K_{\lambda}\left(  u\left(  x,y\right)  \right)  \right)
\end{equation}
where $\lambda=\frac{1}{3}\sqrt{12\mathbf{\nu}^{2}-9\left(  \mu_{\left(
z\right)  }^{2}+\mu_{\left(  \phi\right)  }^{2}\right)  }$ and $u\left(
x,y\right)  =\exp\left(  \frac{\sqrt{3}}{3}\left(  2y+x\right)  \right)\,.  $

One can also consider the Lie algebra $\left\{  K_{\left(  z\right)
},K_{\left(  x\right)  }-\frac{1}{2}K_{\left(  y\right)  },R_{\left(
xz\right)  }-\frac{1}{2}R_{\left(  yz\right)  }\right\}  $ for which we have
the invariant solution
\begin{equation}
\Psi_{2}\left(  x,y,z,\phi\right)  =\left(  \Psi_{3}e^{c_{1}\phi}+\Psi
_{4}e^{-c_{1}\phi}\right)  \left(  \Psi_{1}I_{ic_{1}}\left(  u\left(
x,y\right)  \right)  +\Psi_{2}K_{ic_{1}}\left(  u\left(  x,y\right)  \right)
\right)
\end{equation}
whereas, for the Lie algebra \ $\left\{  K_{\left(  \phi\right)  },K_{\left(
x\right)  }-\frac{1}{2}K_{\left(  y\right)  },~R_{\left(  x\phi\right)
}-\frac{1}{2}R_{\left(  y\phi\right)  }\right\}  $,  the invariant solution is%
\begin{equation}
\Psi_{3}\left(  x,y,z,\phi\right)  =\left(  \Psi_{3}e^{c_{1}z}+\Psi
_{4}e^{-c_{1}z}\right)  \left(  \Psi_{1}I_{ic_{1}}\left(  u\left(  x,y\right)
\right)  +\Psi_{2}K_{ic_{1}}\left(  u\left(  x,y\right)  \right)  \right)
\mathbf{.}%
\end{equation}
The Lie algebras $\left\{  R_{\left(  z\phi\right)  },K_{\left(  x\right)
}-\frac{1}{2}K_{\left(  y\right)  },~R_{\left(  x\phi\right)  }-\frac{1}%
{2}R_{\left(  y\phi\right)  }\right\}  ~$and $\left\{  R_{\left(
z\phi\right)  },K_{\left(  x\right)  }-\frac{1}{2}K_{\left(  y\right)
},~R_{\left(  xz\right)  }-\frac{1}{2}R_{\left(  yz\right)  }\right\}  $
provide the solution of the form%
\begin{equation}
\Psi_{4}\left(  x,y,z,\phi\right)  =\Psi_{1}I_{0}\left(  u\left(  x,y\right)
\right)  +\Psi_{2}K_{0}\left(  u\left(  x,y\right)  \right)  \mathbf{.}%
\end{equation}
In WKB approximation, where the WDW equation (\ref{BSF.20}) reduces to the 
Hamilton-Jacobi equation,  we apply the same approach  as for the case of
General Relativity, Sec. \ref{bianchiIIwkb}, hence we will omit it.
However, we would like to note that the solution for the kination scalar field
is $\phi\left(  t\right)  =c_{\phi}t$, where $c_{\phi}$ is a constant.
In the following section, we discuss how these solutions can be applied, under
a conformal transformation, in the case of $f\left(  \mathcal{R}\right)  $
Hybrid Gravity. This means that the approach can be easily extended to higher-order gravity and non-minimally coupled cases.

\section{Hybrid Gravity in Bianchi cosmology}

\label{hybrid}

In this section we consider the action of the Hybrid metric-Palatini Gravity
with the action of the form \cite{Harko:2011nh,Capozziello:2013uya,Capoz}
\begin{equation}
S=\frac{1}{2\kappa^{2}}\int\mathrm{d}^{4}x\sqrt{-g}[R+f(\mathcal{R}%
)]\mathbf{,}\label{action}%
\end{equation}
where $R$ is the metric Ricci curvature scalar and $f(R)$ is a function of the
Palatini curvature scalar which is constructed by an independent connection
$\hat{\Gamma}$. A variation of the above action with respect to the metric
gives the gravitational field equations
\begin{equation}
G_{\mu\nu}+f^{\prime}(\mathcal{R})\mathcal{R}_{\mu\nu}-\frac{1}{2}%
f(\mathcal{R})g_{ij}=0\label{eq_hyb}%
\end{equation}
where $G_{\mu\nu}$ is the Einstein tensor for metric $g_{ij}$ while $R_{\mu
\nu}$ is the Ricci tensor constructed by the conformally related metric
$h_{\mu\nu}=f^{\prime}(R)g_{\mu\nu}$ \cite{report}. It is well known that
Hybrid Gravity is equivalent to a non-minimally coupled  scalar tensor theory \cite{Capoz}. In
particular, if we consider a new scalar field $\psi=f^{\prime}\left(
\mathcal{R}\right)  $ by using a Lagrange multiplier and the relation between
the two Ricci scalars $R$ and ${\cal R}$, action (\ref{action}) can be written in the
following form
\begin{equation}
S=\frac{1}{2\kappa^{2}}\int\mathrm{d}^{4}x\sqrt{-g}[(1+\psi)R+\frac{3}{2\psi
}\partial^{\mu}\psi\partial_{\mu}\psi-V(\psi)].\label{lagr0}%
\end{equation}
where
\begin{equation}
V(\psi)=\mathcal{R}f^{\prime}(\mathcal{R})-f(\mathcal{R})\text{. }\label{lagr}%
\end{equation}
is a Clairaut equation with the singular solution
\begin{equation}
\frac{\partial V\left(  f^{\prime}\left(  \mathcal{R}\right)  \right)
}{\partial f^{\prime}\left(  \mathcal{R}\right)  }-\mathcal{R}=0\mathbf{.}%
\label{cleq}%
\end{equation}
Furthermore, from the action (\ref{lagr0}) and for the Bianchi spacetimes
(\ref{B2}), we have that the Lagrangian of the field equations is:%

\begin{equation}
L_{f(\mathcal{R})}=N(t)e^{3\lambda}\left[  \left(  1+\psi\right)  \left(
6\dot{\lambda}^{2}-\frac{3}{2}\left(  \dot{\beta}_{1}^{2}+\dot{\beta}_{2}%
^{2}\right)  \right)  +6\dot{\lambda}\dot{\psi}+\frac{3}{2\psi}\dot{\psi}%
^{2}\right]  +\frac{e^{3\lambda}}{N(t)}\left(  V\left(  \psi\right)  +\left(
1+\psi\right)  R^{\ast}\right)  .\label{lagr01}%
\end{equation}
As it is  shown in \cite{Hybrid},  the action (\ref{lagr0}) of the Hybrid Gravity is
equivalent to a phantom minimally coupled scalar field under the conformal
transformation $\bar{g}_{ij}=\left(  1+\psi\right)  g_{ij},$ and action
(\ref{lagr0}) becomes
\begin{equation}
S=\int d^{4}x\sqrt{-g}\left[  \bar{R}+\frac{3}{2}\left(  \frac{2\phi
+3}{2\left(  1+\phi\right)  \phi}\right)  g^{ij}\psi_{;i}\psi_{;j}-\frac
{1}{\left(  1+\psi\right)  ^{2}}V(\psi)\right] \label{HG.03a}%
\end{equation}
where $\bar{R}$ is the Ricci scalar of the conformal metric $\bar{g}_{ij}$;
therefore under the transformation $d\phi=i\sqrt{\left(  \frac{6\psi
+9}{2\left(  1+\psi\right)  \psi}\right)  }d\psi$ and $\bar{V}\left(
\phi\right)  =-\frac{1}{\left(  1+\psi\right)  ^{2}}V(\psi)$,  we have the
action of the form of (\ref{B15}). From the transformation $\psi
\rightarrow\phi$ ,  we  find that
the only potential which admits Lie point symmetries has the following form%
\begin{equation}
V\left(  \psi\right)  =V_{0}\left(  1+\psi\right)  ^{2}\exp\left(
-\kappa\arctan\sqrt{\psi}\right)
\end{equation}
which is the exponential potential in the case of minimally coupled scalar
field cosmology with $\kappa=\kappa\left(  \mu\right)  $ and $\kappa\left(
0\right)  =0,~$i.e. $V\left(  \psi\right)  =V_{0}\left(  1+\psi\right)  ^{2}$. 
This potential transforms to  the constant potential in the case of minimally coupled scalar field \cite{alma}.

\section{Discussion and Conclusions}
\label{conclusion}

In this work we studied the Lie symmetries of the WDW equation in the 
Bianchi Class A spacetimes for General Relativity and  scalar field
cosmologies, considering minimally coupled scalar tensor gravity and non-minimally coupled gravity coming from Hybrid Gravity.
 We applied the Lie invariants in order to construct solutions of
the WDW equation. In the case of General Relativity,  we found exact solutions
of the WDW equation for the Bianchi I and Bianchi II spacetimes. In scalar
field cosmology, we applied the Lie symmetries as a criterion for the
selection of the unknown potential of the scalar field and we were able to
construct exact solutions for the Bianchi I spacetime for zero potential
$V\left(  \phi\right)  =0$, constant potential $V\left(  \phi\right)  =V_{0},$
and exponential potential $V\left(  \phi\right)  =e^{\mu\phi}$. For the
Bianchi II spacetime we obtained solutions for the zero potential case.  In each case, we show that when the
WDW equation is invariant under the action of the three dimensional Lie
algebra with zero commutators, the Hamilton-Jacobi equation of the
Hamiltonian system which was defined by the field equations, can be solved by
the method of separation of variables; that means that the field equations are
Liouville integrable. It is important to note that, in the case of FLRW
scalar cosmology,  we have more potentials where the WDW admits Lie
symmetries.  However,  since the Lie symmetries are connected to the conformal
algebra of the minisuperspace, in the case of FLRW scalar field cosmology,  the
dimension of the minisuperspace is two, which means that the last admits an
infinite number of conformal killing vectors, whereas, for the Bianchi models,
the minisuperspace has dimension four and admits a fifteen dimensional
conformal algebra, i.e. less possible generators for the Lie symmetries of the
WDW equation.

Finally, we studied the case of the Hybrid Gravity in the  Bianchi Class A
spacetimes. Since the Hybrid Gravity is equivalent to a scalar tensor theory,
we were able to related all the potentials we found in the case of minimally
coupled scalar field to that of Hybrid Gravity.

This analysis is important in the sense that can be used in order to construct
solutions of the wave function of the Universe and, at the same time,
conservation laws, and classical solutions for the field equations.  Following the discussion in  \cite{odintsov},  the presence of symmetries gives rise to a straightforward interpretation of  the Hartle criterion: the  symmetries generates oscillatory behaviors in the wave function of the Universe and then allow correlations among physical variables. This give rise to classically observable cosmological solutions. Here we generalized this result considering Bianchi models.  On the other hand,  other general selection rules can be identified in Quantum cosmology,  as discussed in  \cite{kamen1}. This will be the topic of forthcoming papers.

\begin{acknowledgments}
AP acknowledges financial support of FONDECYT grant no. 3160121. 
AW acknowledges financial support from 1445/M/IFT/15 and  DEC-2013/09/B/ST2/03455. SC acknowledges the support of INFN (\textit{iniziative
specifiche} QGSKY and TEONGRAV).
\end{acknowledgments}

\appendix

\section{Symmetries of differential equations and invariant functions}

\label{basictools}

In this appendix,  we briefly discuss the basic properties and definitions
of the symmetries of differential equations (DE). Furthermore we discuss the
application of the Lie symmetries of the WDW equation  showing  that if
the WDW equation admits Lie symmetries, which form a Lie algebra with zero
commutators, then the WDW equation admits oscillatory terms in the solution as
many as the dimension of the minisuperspace and, at the same time, the
Hamiltonian system which defined by the field equations is Liouville
integrable; that is, the field equations can be solved by quadratures.
In such a case, classical cosmological solutions can be derived.

\subsection{Lie point symmetries and invariant functions}

A second order DE is a function $H=H(x^{i},u^{A},u_{,i}^{A},u_{,ij}^{A})$ in
the jet space $B_{M}$, where $x^{i}$ are the independent variables and $u^{A}$
are the dependent ones. Let
\begin{equation}
\mathbf{X}=\xi^{i}(x^{k},u^{B})\partial_{x^{i}}+\eta^{A}(x^{k},u^{B}%
)\partial_{u^{A}}.\label{pr.03}%
\end{equation}
be the generator of the infinitesimal point transformation
\begin{align}
\bar{x}^{i} &  =x^{i}+\varepsilon\xi^{i}(x^{k},u^{B})\label{pr.01}\\
\bar{u}^{A} &  =\bar{u}^{A}+\varepsilon\eta^{A}(x^{k},u^{B})\mathbf{.}%
\label{pr.02}%
\end{align}

The function $H=0,$ is invariant under the action of the infinitesimal point
transformation (\ref{pr.01}),(\ref{pr.02}) if there exists a function
$\lambda$ such that \cite{Lie67a}
\begin{equation}
\mathbf{X}^{[2]}(H)=\lambda H~\label{pr.04}%
\end{equation}
The vector field $X$ is called Lie point symmetry of the function $H$ and
$X^{[n]}~$is the second prolongation of $X$ in the jet space $B_{M}$
\begin{equation}
\mathbf{X}^{[2]}=\mathbf{X}+\eta_{i}^{A}\partial_{\dot{x}^{i}}+\eta_{ij}%
^{A}\partial_{u_{ij}^{A}}\mathbf{,}\label{pr.05}%
\end{equation}
where%
\begin{equation}
\eta_{i}^{A}=\eta_{,i}^{A}+u_{,i}^{B}\eta_{,B}^{A}-\xi_{,i}^{j}u_{,j}%
^{A}-u_{,i}^{A}u_{,j}^{B}\xi_{,B}^{j}~,\label{pr.06}%
\end{equation}
and
\begin{align}
\eta_{ij}^{A} &  =\eta_{,ij}^{A}+2\eta_{,B(i}^{A}u_{,j)}^{B}-\xi_{,ij}%
^{k}u_{,k}^{A}+\eta_{,BC}^{A}u_{,i}^{B}u_{,j}^{C}-2\xi_{,(i|B|}^{k}u_{j)}%
^{B}u_{,k}^{A}\nonumber\\
&  -\xi_{,BC}^{k}u_{,i}^{B}u_{,j}^{C}u_{,k}^{A}+\eta_{,B}^{A}u_{,ij}^{B}%
-2\xi_{,(j}^{k}u_{,i)k}^{A}-\xi_{,B}^{k}\left(  u_{,k}^{A}u_{,ij}^{B}%
+2u_{(,j}^{B}u_{,i)k}^{A}\right)  \mathbf{.}\label{pr.07}%
\end{align}

The importance of Lie point symmetries is that the last one can be used in
order to reduce an order of a differential equation. When a reduction is
possible, one can determine invariant solutions or transform them to another
ones \cite{Bluman}. From condition (\ref{pr.04}) one defines the Lagrange
system%
\[
\frac{dx^{i}}{\xi^{i}}=\frac{du}{\eta}=\frac{du_{i}}{\eta_{\left[  i\right]
}}=\frac{du_{ij}}{\eta_{ij}}%
\]
whose solution provides the characteristic functions
\[
Z^{\left[  0\right]  }\left(  x^{k},u\right)  ,~Z^{\left[  1\right]  i}\left(
x^{k},u,u_{i}\right)  ,Z^{\left[  2\right]  }\left(  x^{k},u,u_{,i}%
,u_{ij}\right)  .
\]
The solution $Z^{\left[  k\right]  }$ is called the $k$th order invariant of
the Lie symmetry vector (\ref{pr.03}). By writing the DE in terms of the
invariants $Z^{\left[  k\right]  }$, we can reduce the order of the DE, for
details see for instance \cite{Stephani,Bluman}. Below we discuss the
application of the Lie symmetries and of the Lie invariants for the WDW equation.

\subsection{Reduction and invariant solutions of the WDW equation by Lie point
symmetries}

In order to determine the Lie symmetries of the WDW equation we apply a
geometric method which is established by Paliathanasis \& Tsamparlis
\cite{AnIJGMMP}. The method relates the Lie symmetries of the Klein Gordon
equation to the conformal algebra of the underlying geometry. Hence, in the
following we will not present the construction of the Lie symmetries of the
WDW equation but we will give the results.

In particular, the general form of a Lie symmetry of the WDW equation is
\begin{equation}
\mathbf{X}=\xi^{i}\left(  x^{k}\right)  \partial_{i}+\left(  \frac{2-n}{2}%
\psi\Psi+a_{0}\Psi\right)  \partial_{\Psi}\mathbf{,}\label{CJ.8}%
\end{equation}
where $\xi^{i}\left(  x^{k}\right)  $ is a CKV of the metric which defines the
conformal Laplace operator, (in our consideration the minisuperspace) and
$\psi\left(  x^{k}\right)  $ is the conformal factor of the CKV, recall that
since~$\xi^{i}$ is a CKV of $g_{ij}$, it means that, $L_{\xi}g_{ij}=2\psi
g_{ij}$.

Furthermore, it is possible to consider a coordinate transformation
$x^{i}\rightarrow\bar{x}^{i}$ so that $\xi^{i}\left(  x^{k}\right)
\partial_{i}\rightarrow\partial_{J}~$ (these are called normal coordinates).
In the normal coordinates the symmetry vector takes the following simple form, %

\begin{equation}
\mathbf{X}=\partial_{J}+\left(  \frac{2-n}{2}\psi\Psi+a_{0}\Psi\right)
\partial_{u}\label{CJ.8a}%
\end{equation}
where now either with the method of Lie invariants, or with the method of
linear differential operators \ (see \cite{bar1} for details) we find the
following expression for the solution of $\Psi$,
\begin{equation}
\Psi\left(  \bar{x}^{b},\bar{x}^{J}\right)  =\Phi\left(  \bar{x}^{b}\right)
\exp\left[  \int\left( \frac{2-n}{2}\psi-Q_{0}\right)  d\bar{x}^{J}\right]
\label{CJ.13}%
\end{equation}
from which follows again that the coordinate $x^{J}$ is factored out \ from
the solution $\Psi.$

Furthermore, in \cite{AnIJGMMP} it was also shown that the symmetries of the
WDW equation can be used in order to find Noether symmetries for classical
particles\footnote{A similar methodology by using the variatonal symmetries
has been established recently in \cite{Chr1}}. The exact relation among the
Lie symmetries of the WDW equation, the Noetherian conservation laws of the
field equations and the interpretability conditions are given in \cite{bar1}.

\end{document}